\documentclass[superscriptaddress,prd,aps,amssymb,nofootinbib,twocolumn]{revtex4}
\usepackage{graphicx, wrapfig}

\newcommand{\beq}{\begin{equation}}
\newcommand{\eeq}{\end{equation}}
\newcommand{\beqn}{\begin{eqnarray}}
\newcommand{\eeqn}{\end{eqnarray}}

\newcommand{\zD}{{\raise1.0ex\hbox{${}^{\ \circ}$}}\!\!\!\!\!D}
\newcommand{\alone}{{\raise0.5ex\hbox{${}^{\ 1}$}}\!\!\!\!\alpha}
\newcommand{\Od}{{O}}

\newcommand{\dl}{\delta}

\newcommand{\nalam}{\mathrel{\raise0.9ex\hbox{$^\lambda$}\mkern-14mu
\lower0.0ex\hbox{$\nabla$}}}

\newcommand{\be}{\begin{equation}}
\newcommand{\ee}{\end{equation}}
\newcommand{\bea}{\begin{eqnarray}}
\newcommand{\eea}{\end{eqnarray}}
\newcommand{\beaa}{\begin{eqnarray*}}
\newcommand{\eeaa}{\end{eqnarray*}}
\newcommand{\ben}{\begin{enumerate}}
\newcommand{\een}{\end{enumerate}}

\newcommand{\bmm}{\bar{m}}

\newcommand{\bF}{\bar{F}}
\newcommand{\hE}{\widehat{E}}

\newcommand{\ba}{\bar{a}}
\newcommand{\bv}{\bar{v}}

\newcommand{\bgamma}{{\bar{\gamma}}}

\newcommand{\Fone}{{F_{\,\rm I}}}
\newcommand{\bFone}{{\bar F_{\,\rm I}}}
\newcommand{\FPN}{{F_{\rm PN}}}

\newcommand{\FSPN}{{F_{\rm SPN}}}

\newcommand{\hEPM}{\widehat{E}_{\rm PM}}
\newcommand{\EPM}{{E_{\rm PM}}}
\newcommand{\eone}{{e_{\rm I}}}
\newcommand{\ePN}{{e_{\rm PN}}}
\newcommand{\eSPN}{{e_{\rm SPN}}}
\newcommand{\LPM}{{L_{\rm PM}}}
\newcommand{\lone}{{\ell_{\rm I}}}
\newcommand{\lPN}{{\ell_{\rm PN}}}
\newcommand{\lSPN}{{\ell_{\rm SPN}}}

\usepackage{epsfig}

\begin{document}
\preprint{}

%
%

\title{Circular solution of two unequal mass particles in
post-Minkowski approximation}
\author{Matthew M. Glenz}
\affiliation{
Department of Physics, University of Wisconsin-Milwaukee, P.O. Box 413,
Milwaukee, WI 53201}
\author{K\=oji Ury\=u}
\affiliation{
Department of Physics, University of Wisconsin-Milwaukee, P.O. Box 413,
Milwaukee, WI 53201}
\affiliation{
Department of Physics, University of the Ryukyus,
1 Senbaru, Nishihara, Okinawa 903-0213 Japan}
%
%

\begin{abstract}
A Fokker action for post-Minkowski approximation with the first
post-Newtonian correction is introduced in our previous paper,
and a solution for the helically symmetric circular orbit is
obtained. We present supplemental results for the circular solution
of two unequal mass point-particles.  Circular solutions for
selected mass ratios are found numerically, and analytic formulas
in the extreme mass ratio limit are derived. The leading terms of
the analytic formulas agree with the first post-Newtonian formulas
in this limit.
\end{abstract}


\maketitle

\section{Introduction}
\label{sec:intro}

In our previous paper \cite{FU06} (Paper I), a Fokker action that
describes a time symmetric interaction of point-particles is
introduced in the framework of a post-Minkowski approximation.
From two types of Fokker
action, a parametrization invariant action with a post-Newtonian
correction and an affinely parametrized action, the equations of
motion and expressions for conserved energy and angular momentum
are derived following the variational calculation of Ref. \cite{ds54}. We
find a solution describing a helically symmetric circular orbit in
the post-Minkowski approximation (with post-Newtonian corrections)
that is analogous
to the circular solution of two charges obtained by Schild for the
electromagnetic interaction \cite{sc63}. We report here results
supplementing those of Paper I: numerically computed solution
sequences for unequal mass particles, and analytic formulas in the
extreme mass ratio limit.  The latter results agree with the first
post-Newtonian (1PN) formulas; hence a consistency of our model is
confirmed in this limit.

\section{Formulas for circular solutions}

We present a set of formulas governing the helically symmetric circular
orbits of two point particles, $\{m,v\}$ and $\{\bmm,\bv\}$, and
derive analytic expressions in the extreme mass ratio limit $q:=m/\bmm
\rightarrow 0$. The set of algebraic equations is solved numerically
for a fixed binary separation to specify each circular orbit. The
result for the unequal mass binary orbit is presented in Sec.
\ref{sec:numerical}. Units of $G=c=1$ are used in this report.

\subsection{Parametrization invariant model with post-Newtonian correction}

\subsubsection{Circular solution}
\label{sec:pinveqs}

We first list the result from Paper I for the parametrization
invariant model with 1PN correction
terms. The (integrated) equations of motion for particles $m$ and
$\bmm$ are written in terms of the velocities, $v$ and $\bv$, of
particles $m$ and $\bmm$ which are related to the orbital radius
by $a:= v/\Omega$ and $\ba:= \bv/\Omega$,
\beqn
-m\gamma^2 v\Omega
&=& - m\bmm\gamma^2\bgamma \Omega^2
\big[\,F(\varphi,v,\bv)
\nonumber\\
&&+(m+\bmm)\Omega\, \Fone (\varphi,v,\bv,\gamma,\bgamma)\,
\big],
\label{eq:eompn}
\\[2mm]
-\bmm\bgamma^2 \bv\Omega
&=& - m\bmm\gamma\bgamma^2 \Omega^2
\big[\,\bF(\varphi,v,\bv)
\nonumber\\
&&+(m+\bmm)\Omega\, \bFone (\varphi,v,\bv,\gamma,\bgamma)\,
\big].
\label{eq:eompnbar}
\eeqn
As shown below, $\{\varphi,v,\bv,\gamma,\bgamma\}$ are not
independent.
The functions $F(\varphi,\bv,v)=\bF(\varphi,v,\bv)$ are
the post-Minkowski terms, while $\Fone(\varphi,\bv,v,\bgamma,\gamma)
= \bFone (\varphi,v,\bv,\gamma,\bgamma)$ is either of two alternative
1PN
correction terms that agree at 1PN order:
$\Fone = \FPN(\varphi,v,\bv,\gamma,\bgamma)$ derived from
a non-relativistic correction, or $\Fone =
\FSPN(\varphi,v,\bv,\gamma,\bgamma)$ derived from a special
relativistically invariant correction.
%
\beqn
&&F(\varphi,v,\bv) \,:=\,
- 4 \frac1{(\varphi+v\bv\sin\varphi)^2}
\bigg\{(1+v\bv\cos\varphi)\bv
\nonumber\\
&&
\times
(\varphi\cos\varphi-v^2\sin\varphi)
+\frac12 v(1-\bv^2)(\varphi+v\bv\sin\varphi)
\nonumber\\
&&- \frac12\big[\bv\sin\varphi(\varphi+v\bv\sin\varphi)
+ (1+v\bv\cos\varphi)(v+\bv\cos\varphi)
\nonumber\\
&&- \frac{v}{1-v^2}(\varphi+v\bv\sin\varphi)^2
\big]\Phi(\varphi,v,\bv)
\bigg\},
\\[2mm]
&&\FPN(\varphi,v,\bv,\gamma,\bgamma) \,:=\,
- \frac{1}{\gamma^2 \bgamma^2 (v+\bv)^{3}}
\left[ 1+\frac12\gamma ^2 v(v+\bv)\right],
\nonumber\\
\\
&&\FSPN(\varphi,v,\bv,\gamma,\bgamma) \,:=\,
-\frac1{(\gamma \bgamma)^{5/2}}
\frac{1}{\left( \varphi +v\bv\sin \varphi
\right)^2}
\nonumber\\
&&
\times
\bigg\{ \frac34\gamma ^{2}v
+\frac{\bv\sin \varphi }
{\varphi +v\bv\sin \varphi }+\frac{\left( 1+v\bv\cos \varphi \right)
\left( v+\bv\cos \varphi \right) }{\left( \varphi +v\bv\sin \varphi
\right)^2}\bigg\}.
\nonumber\\
\eeqn
The function $\Phi(\varphi,v,\bv)$ is defined by \beq
\Phi(\varphi,v,\bv) \,:=\,\frac{(1+v\bv\cos\varphi)^2 -
\frac12(1-v^2)(1-\bv^2)} {\varphi+v\bv\sin\varphi}.
\label{eq:fncPhi} \eeq
For the parametrization invariant models, $\gamma$ and $\bgamma$
are derived from a flat-space normalization of the four-velocity, \beq \gamma =
(1-v^2)^{-\frac12},\ \ \ \ \ \ \bgamma = (1-\bv^2)^{-\frac12}. \eeq
The retarded angle $\varphi$ is the positive root of $\varphi^2 =
v^2 + \bv^2 +2v \bv \cos\varphi$.


\subsubsection{Extreme mass ratio limit}

The extreme mass ratio limit $q:=m/\bmm
\rightarrow 0$ is identical to the limit $\bv\rightarrow
0$ with $\Omega$ fixed. In the limit $\bv\rightarrow 0$, we may assume that
$v$ and $\bmm$ remain finite. Consequently, we have
$\bgamma \rightarrow 1$,
$\varphi \rightarrow v$, and $\bmm \rightarrow M$, where
$M:=m+\bmm$ is the total mass. With $v$ and $\Omega$ regarded as
independent variables, Eq.~(\ref{eq:eompn}) is a
quadratic equation for $\Omega M$, whose $q=0$ form is
\beq
\Fone\,(\Omega M)^2+
F\,(\Omega M)-v=0,
\eeq
with physical solution
\beq
\Omega M = \frac1{2\Fone}\left(-F+\sqrt{F^2+4\Fone
v}\right). \label{eq:omesol}
\eeq
The functions $F$ (the post-Minkowski term), $\Fone=\FPN$ and
$\Fone=\FSPN$ (the alternative forms of the 1PN correction)
for $q=0$
become \beqn &&\!\!\!\!\!\!\!
F(\varphi,v,\bv)
\,=\,
\frac{1-3v^2}{v^2(1-v^2)},
\\[2mm]
&&\!\!\!\!\!\!\!
\FPN(\varphi,v,\bv,\gamma,\bgamma)
\,=\,
- \frac1{v^{3}}\left(1-\frac12 v^2\right),
\\[2mm]
&&\!\!\!\!\!\!\!
\FSPN(\varphi,v,\bv,\gamma,\bgamma)
\,=\,
-\frac{(1-v^2)^{1/4}}{v^3}
\left(1-\frac14 v^2 \right),
\eeqn
where
$\Phi$ has the form
$ \Phi(\varphi,v,\bv)
= (1+v^2)/(2\,v)$.

Note that the parametrization invariant post-Minkowski model is
derived by setting $\Fone=0$, and therefore $\Omega M = v/F$. In the
$q \rightarrow 0$ limit, this is written
$ \Omega M = v^3(1-v^2)/(1-3v^2). $

\subsubsection{Energy and angular momentum formulas}

The conserved energy and angular momentum for the parametrization
invariant model are written
\beq
E \,=\, \EPM + \eone, \quad {\rm and }\quad
L \,=\, \LPM + \lone,
\eeq
where $\EPM$ and $\LPM$ are the post-Minkowski terms
\beqn
\EPM &=& \frac{m}{\gamma} + \frac{\bmm}{\bgamma}
\\
\LPM &=& 2m\bmm\gamma\bgamma\,\Phi(\varphi,v,\bv),
\eeqn
and $\eone$ and $\lone$ are the parametrization invariant 1PN
corrections $\eone = \ePN$ and $\lone=\lPN$, or those of the special
relativistically invariant model $\eone = \eSPN$ and $\lone=\lSPN$
given by \beqn \eone &=& \frac12\Omega \lone, \quad \mbox{(for both
$\eone = \ePN$ and $\eSPN)$,}
\label{epn}\\
\ell_{\rm PN} &=& -\frac{m\bar m(m+\bar m)\Omega}{\gamma\bar\gamma(v+\bar v)^2},
\label{lpn}\\
\ell_{\rm SPN} &=& -\frac{m\bar m(m+\bar m)\Omega}{(\gamma\bar\gamma)^{3/2}}
\frac1{(\varphi+v\bar v \sin\varphi)^2}.
\label{lspn}\eeqn

In the $q\rightarrow 0$ limit,
the conserved energy and angular momentum normalized by the mass
remain finite.
Subtracting the mass of the heavier particle from the post-Minkowski
energy, $\hEPM := \EPM - \bmm$,
and taking the limit $\bv\rightarrow 0$ with $\bmm\rightarrow M$,
we have
\beqn
&&
\frac{\hEPM}{m} \,=\, m(1-v^2)^{1/2},
\\[2mm]
&&
\frac{\LPM}{mM} \,=\,  \frac{1+v^2}{v(1-v^2)^{1/2}},
\\[2mm]
&&
\frac{\ell_{\rm PN}}{mM} \,=\, - \frac{(1-v^2)^{1/2}}{v^2}\Omega M,
\label{lpn}
\\[2mm]
&&
\frac{\ell_{\rm SPN}}{mM} \,=\, - \frac{(1-v^2)^{3/4}}{v^2}\Omega M.
\label{lspn}
\eeqn

\subsubsection{Solution sequence in $q \rightarrow 0$ limit}

In Paper I, it is proved that the first law of thermodynamics that
relates the changes in the conserved energy and the angular
momentum, $dE = \Omega dL$, is satisfied by binary solutions
derived from the parametrization invariant Fokker action. This
relation is used to cross check the analytic formula in the
$q\rightarrow 0$ limit above as well as the numerical solutions
shown in the next section by calculating
${d\hE}/{dv}=\Omega{dL}/{dv}$, where $\hE:=\hEPM + \eone$.

In the parametrization invariant post-Minkowski model, the normalized
angular velocity of a particle $m$, $\Omega M$, is defined in an
interval $0\le v < 1/\sqrt{3}$, and it becomes infinite at
$v=1/\sqrt{3}$. With the 1PN correction $\Fone=\FPN$, the range of
finite $\Omega M$ is approximately $0\le v \alt 0.361598$, and with
the special relativistic invariant 1PN correction $\Fone=\FSPN$, it
is $0\le v \alt 0.36166$.   Newtonian point particles have no innermost
stable circular orbit (ISCO), but adding a 1PN correction
to the Newtonian orbit recovers the ISCO that is present in the exact
theory.  In the post-Minkowski framework, we find that the
existence of an ISCO depends on our choice among actions that are
equivalent to first post-Minkowski order. In particular, we will see
that the parametrization-invariant action leads to sequences
with no ISCO {\em even when 1PN terms are included}.
This is plausibly due to the fact that the sequences associated
with the parametrization-invariant action terminate before reaching
the angular velocity of an ISCO.  In fact, in the standard
1PN formalism, an ISCO occurs at an unrealistically high value of
angular velocity, namely $\Omega M = 0.544$, where the 2PN and 3PN
values are $\Omega M = 0.124$ and $0.0867$ for $q=0$, respectively
\cite{Blanc02}.

Curiously, however, as we note in the next section,
sequences associated with the affinely parametrized
action do have an ISCO, in this case at an unrealistically small values
of $\Omega M$.

Finally, we show that these results of post-Minkowski plus 1PN
corrections agree with the 1PN formula in the $q\rightarrow 0$
limit. In Eq.~(\ref{eq:omesol}), an expansion of $\Omega M$ in the
small $v$ limit becomes $\Omega M = v^3 + 3 v^5 + \Od(v^7)$ for both
PN and SPN models, and this is inverted to write $v$ in terms of
small $\Omega M$ as
$
v = (\Omega M)^{1/3} -  \Omega M
+\Od\big((\Omega M)^{5/3}\big).
$
Substituting this into the
energy and angular momentum formulas, the leading two terms agree
with the post-Newtonian formulas (see e.g. \cite{Blanc02}) up to the
1PN order for the extreme mass ratio $q\rightarrow 0$,
\beqn
\frac{\hE}{m} &=& -\frac12(\Omega M)^{2/3} +  \frac38(\Omega
M)^{4/3} +\Od\big((\Omega M)^{2}\big),
\\[2mm]
\frac{L}{mM} \!\!&=& \!\!
\frac1{(\Omega M)^{1/3}}\left[1+\frac32(\Omega M)^{2/3}
+\Od\big((\Omega M)^{4/3}\big) \right],
\eeqn

\subsection{Affinely parametrized model}

\subsubsection{Circular solution}
\label{sec:affineqs}

For the affinely parametrized post-Minkowski model, analogous forms
of Eqs.~(\ref{eq:eompn}) and (\ref{eq:eompnbar}) are written \beqn
-m\gamma^2 v\Omega &=& - m\bmm\gamma^2\bgamma \Omega^2
F^A(\varphi,v,\bv), \label{eq:eomaf}
\\[2mm]
-\bmm\bgamma^2 \bv\Omega
&=& - m\bmm\gamma\bgamma^2 \Omega^2  \bF^A(\varphi,v,\bv),
\label{eq:eombaraf}
\eeqn
where the function $F^A(\varphi,\bv,v) = \bF^A(\varphi,v,\bv)$ is
written \beqn && F^A(\varphi,v,\bv) \,:=\, - 4
\frac1{(\varphi+v\bv\sin\varphi)^2}
\nonumber\\
&&
\times\bigg\{(1+v\bv\cos\varphi)\bv(\varphi\cos\varphi-v^2\sin\varphi)
\nonumber\\
&&
+\frac12 v(1-\bv^2)(\varphi+v\bv\sin\varphi)
- \frac12\big[\bv\sin\varphi(\varphi+v\bv\sin\varphi)
\nonumber\\
&&
+ (1+v\bv\cos\varphi)(v+\bv\cos\varphi)
\big]\Phi(\varphi,v,\bv)
\bigg\}.
\eeqn
For the affinely parametrized world line, $\gamma$ and $\bgamma$
satisfy
\beqn
-\gamma^2(1-v^2) + 4\bmm\gamma^2\bgamma\Omega\,\Phi(\varphi,v,\bv)
= -1,
\label{eq:norm1}
\\[2mm]
-\bgamma^2(1-\bv^2) + 4m\gamma\bgamma^2\Omega\,\Phi(\varphi,v,\bv)
= -1.
\label{eq:norm2}
\eeqn
%

In the limit of $q\rightarrow 0$ (or more directly $\bv\rightarrow
0$),
$ F^A(\varphi,v,\bv)
=(1-v^2)/v^2.
$
%
From Eq.~(\ref{eq:eomaf}) and (\ref{eq:norm1}), we have
$ \gamma =
(1-v^2)^{1/2}/(1-4v^2-v^4)^{1/2},
$
while
in Eq.~(\ref{eq:norm2}), taking $\bv\rightarrow 0$ and $m
\rightarrow 0$ yields $\bgamma \rightarrow 1$. As a result we have
in the extreme mass ratio $q \rightarrow 0$,
\beq
\Omega M
\,=\,
\frac{v^3}{1-v^2}.
\eeq
%
%
%
\begin{figure}[t]
\includegraphics[scale=1.4,clip]{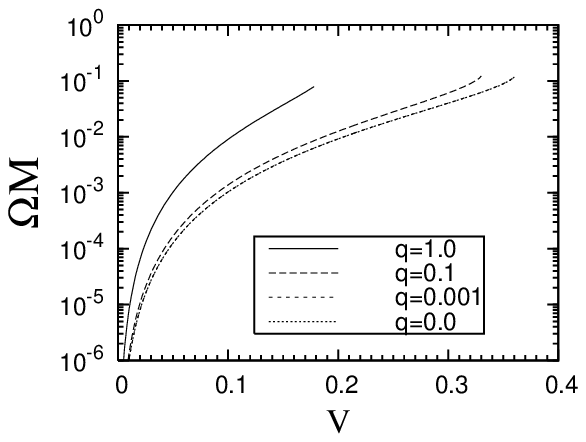}
\caption{\label{fig:SPNOV} Angular velocity, in dimensionless form
$\Omega M$, is plotted against the velocity of the lighter particle
for 3 mass ratios and the $q\rightarrow 0$ limit for the
parametrization invariant model with SPN correction. Curves of the
analytic solution for $q \rightarrow 0$ and that of $q=0.001$
overlap each other in the plot.
}
%
\includegraphics[scale=1.4,clip]{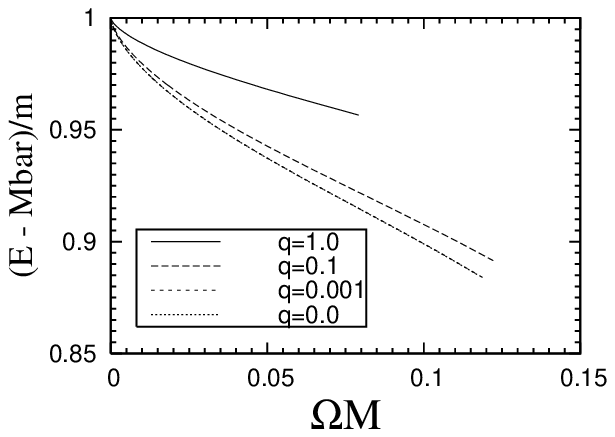}
\caption{\label{fig:SPNEO} Energy, in
dimensionless form
$\hat{E}/m$, where $\hat{E}=E-\bar{m}$,
is plotted against $\Omega M$ for the same models as
in Fig.\ref{fig:SPNOV}.
}
\end{figure}

\subsubsection{Energy and angular momentum formula}

The conserved energy and angular momentum for the affinely parametrized
model are written
\beqn
E &=& \frac{m}{\gamma}+\frac{\bmm}{\bgamma}
+ 4m\bmm\gamma\bgamma\Omega\,\Phi(\varphi,v,\bv),
\\
L &=&
2m\bmm\gamma\bgamma\,\Phi(\varphi,v,\bv),
\label{eq:angsolaffine} \eeqn where the form of
$\Phi(\varphi,v,\bv)$ is the same as that of the parametrization
invariant model (\ref{eq:fncPhi}). Using Eq.(\ref{eq:norm1}) and
(\ref{eq:norm2}), the energy can be rewritten \beq E =
\frac12\frac{m}{\gamma} + \frac12 m\gamma(1-v^2)
+\frac12\frac{\bmm}{\bgamma} + \frac12 \bmm\bgamma(1-\bv^2). \eeq

In the $q\rightarrow 0$ limit, the energy without the rest mass of
the heavier particle, $\hE\,:=\, E-\bmm$, and the angular momentum
become
\beqn
&&
\frac{\hE}{m}
\,=\,
\frac{(1-3v^2)}{[(1-v^2)(1-4v^2-v^4)]^{1/2}},
\\[2mm]
&&
\frac{L}{mM}
\,=\,
\frac{1+v^2}{v}\left(\frac{1-v^2}{1-4v^2-v^4}\right)^{1/2}.
\eeqn
%
%

%
%
\begin{figure}[t]
\includegraphics[scale=1.4,clip]{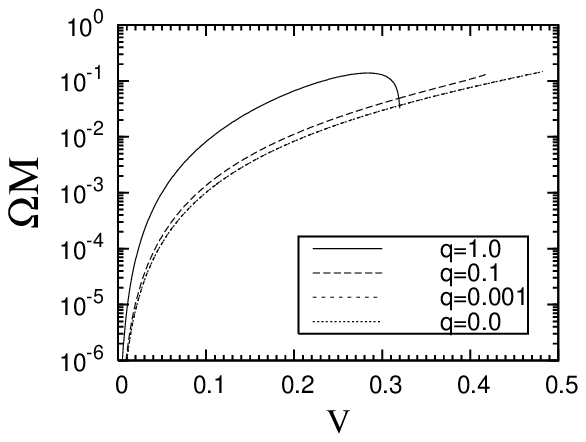}
\caption{\label{fig:AFOV}
Same as Fig.~\ref{fig:SPNOV} but for the affinely parametrized
model. In the $q=1$ case the ISCO occurs at $v\sim$0.184.}
%
%
\includegraphics[scale=1.4,clip]{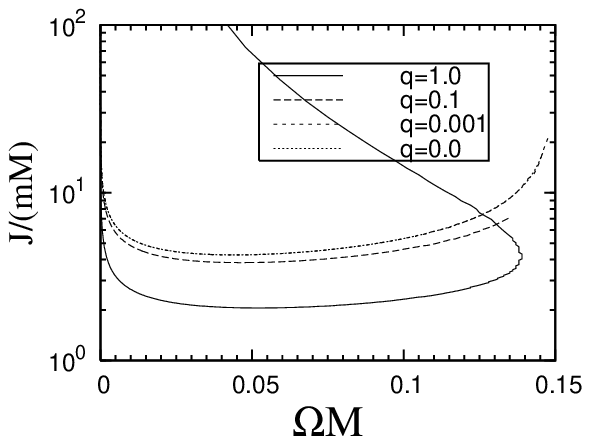}
\caption{\label{fig:AFJO} Angular momentum, in dimensionless form
$J/(mM)$, is plotted against $\Omega M$ for the same models as in
Fig.~\ref{fig:AFOV}. Minima of each curve corresponds to the ISCO.
}
\end{figure}

\subsubsection{Solution sequence in $q \rightarrow 0$ limit}

The first law $\dl E=\Omega\dl L$ is also satisfied for the affinely
parametrized model, and hence one can cross check formulas in the
$q\rightarrow 0$ limit using the relation $d\hE/dv=\Omega dL/dv$.
Although the normalized angular velocity of a particle $m$, $\Omega
M$, is finite in an interval $v\in[0,1)$, the redshift factor
$\gamma$ as well as conserved quantities $E$ and $L$ become infinite
at $v=\sqrt{\sqrt{5}-2}\approx0.485868$, which corresponds to $
\Omega M = {\left(\sqrt{5}-2\right)^{3/2}}/({3-\sqrt{5}})\approx
0.150142. $

In this interval, $v\in\big[0,\sqrt{\sqrt{5}-2}\big)$, the energy
and angular momentum have a simultaneous minima at $
v=\sqrt{({1+2^{4/3}-2^{5/3}})/{3}}\approx 0.339136, $ which
corresponds to $\Omega M \approx 0.0440743$.

\section{Numerical solutions for the unequal mass circular orbit}
\label{sec:numerical}

A circular solution is calculated from algebraic equations given in
Sec.\ref{sec:pinveqs} for the parametrization invariant model, and
\ref{sec:affineqs} for the affinely parametrized model. It turned
out that a convenient way to find a solution is (1) fix the ratio of
velocities $v/\bv$ and determine the corresponding mass ratio from
the equations of motion, then (2) change the velocity ratio to
adjust the value of the mass ratio to a fixed value (using the
bisection method, for example).

In Figs.~\ref{fig:SPNOV} and \ref{fig:SPNEO}, the plots of the
parametrization invariant model with SPN correction terms are
presented for three mass ratios, $q=1.0$, $0.1$, and $0.001$.  Plots
for the case with PN correction terms are not shown here, and they are
qualitatively the same as SPN cases. The analytic solution in the
$q\rightarrow 0$ limit is also plotted and it overlaps with the
$q=0.001$ line in the plots.

In Figs.~\ref{fig:AFOV} and \ref{fig:AFJO}, the plots of the
affinely parametrized model are presented for the same mass ratios
as above.  The solutions of the affinely parametrized model are
markedly different; for any mass ratio $q\in[0,1]$, we found a
simultaneous minima in the energy and angular momentum which
corresponds to the ISCO.

\section{Discussion}

Agreement between the energy and angular momentum formulas of the
1PN circular solution, and those of the parametrization invariant
post-Minkowski model with post-Newtonian correction, is exhibited
only for the extreme mass ratio limit in this report. For an
arbitrary mass ratio one needs to expand the retarded angle
$\varphi$ to the next order in the velocities, $v$ and $\bv$, as
$\varphi \approx (v+\bv)(1-v\bv/2)$, and the rest of the calculation
closely parallels that of the $q = 0$ case.

\begin{acknowledgments}
We thank John L. Friedman for discussions and careful reading of the
manuscript. This work was supported by NSF grants Nos. PHY0071044
and PHY0503366, the Lynde and Harry Bradley Foundation, and the
National Space Grant College and Fellowship Program and the
Wisconsin Space Grant Consortium.
\end{acknowledgments}

\end{document}